\documentstyle[11pt,psfig]{article}
\begin{document}
\renewcommand{\thepage}{ }
\begin{titlepage}
\title{
\hfill
\vspace{1.5cm}
{\center \bf Designability, thermodynamic stability, and dynamics
in protein folding: a lattice model study
}
}
\author{
R\'egis M\'elin$^{(1),(2)}$, Hao Li$^{(2),(3)}$,
Ned S. Wingreen$^{(2)}$,
and Chao Tang$^{(2)}$\\
{}\\
{$^{(1)}$ Centre de Recherches sur les Tr\`es basses
temp\'eratures\thanks{U.P.R. 5001 du CNRS}
(CRTBT-CNRS)}\\
{Laboratoire conventionn\'e avec l'Universit\'e Joseph Fourier}\\
{BP 166X, 38042 Grenoble C\'edex, France}\\
{}\\
$^{(2)}$ NEC Research Institute,
4 Independence Way, Princeton, NJ 08540, USA\\
{}\\
$^{(3)}$ Center for Studies in Physics and Biology,\\
Rockefeller University, 1230 York Avenue, New York,
NY 10021, USA\\
}
\date{}
\maketitle
\begin{abstract}
\normalsize
In the framework of a lattice-model study of protein folding,
we investigate the interplay between designability, thermodynamic
stability, and kinetics. To be ``protein-like'', heteropolymers
must be thermodynamically stable, stable against mutating
the amino-acid sequence, and must be fast folders.
We find two criteria which, together, guarantee that a 
sequence will be ``protein like'': i) the ground state is
a highly designable stucture, {\it i. e.} the native structure is
the ground state of a large number of sequences, and
ii) the sequence has a large $\Delta/\Gamma$ ratio,
$\Delta$ being the average energy separation between the
ground state and the excited compact conformations,
and $\Gamma$ the dispersion in energy of excited compact conformations.
These two criteria are not incompatible since, on average,
sequences whose ground states are highly designable structures
have large $\Delta/\Gamma$ values.
These two criteria require knowledge only of the
compact-state spectrum.
These claims are substantiated by the study of 45 sequences,
with various values of $\Delta/\Gamma$ and various
degrees of designability, by means of a Borst-Kalos-Lebowitz algorithm,
and the Ferrenberg-Swendsen histogram optimization method.
Finally, we report on the reasons for slow folding.
A comparison between a very slow folding sequence,
an average folding one and a fast folding
one suggests that slow folding originates from a  
proliferation of nearly compact low-energy conformations,
not present for fast folders. 
\end{abstract}
\end{titlepage}

\newpage
\renewcommand{\thepage}{\arabic{page}}
\setcounter{page}{1}
\baselineskip=17pt plus 0.2pt minus 0.1pt

\newpage

\section{Introduction}
Among the set of all possible linear amino-acid heteropolymers,
only very few are ``protein like''. For a heteropolymer to be 
``protein like'', three requirements must be met:
(i) The heteropolymer must be thermodynamically stable:
at thermal equilibrium, it
must spend a large fraction of its time in the ground state. 
(Anfissen \cite{Anfissen} has shown the ground state of a protein to be the
biologically active configuration.)
(ii) The heteropolymer must have a fast folding time: 
the native state should be kinetically
accessible, typically within milliseconds to seconds
for real proteins.  (iii) The ``protein-like'' heteropolymer must 
be stable against mutations: if an amino acid
is mutated into another one, the native structure should typically
be preserved. 

Why are some sequences of amino acids ``protein like'' while others
are not? Since theoretical methods cannot yet reliably find the
ground state of real amino-acid chains, we address this question
within a simple lattice model. Lattice models have been 
widely used in the study of protein folding dynamics
\cite{Chan,Shakhnovich0,Shakhnovich1,Thirumalai,Klimov,Socci,Banavar}.
The main ingredients of these lattice models
are (i) the protein is viewed as a heteropolymer on a cubic lattice,
and
(ii) non-covalently-bonded nearest neighbor
monomers experience an interaction
that depends on monomer type.
In one class of lattice models, 
the interactions between adjacent monomers are chosen as random
variables from a continuous probability distribution (see for instance
\cite{Shakhnovich0,Shakhnovich1,Klimov}). We adopt another approach, namely
a so-called H-P model, where the monomers come
in only two types, H(ydrophobic) or P(olar) \cite{Chan,Socci}.
The main physical motivation for studying H-P models is that the 
specific ground-state configuration of real proteins appears to be 
largely determined by optimal burial of hydrophobic amino acids  
away from water\cite{Hydro}. It was also shown \cite{NEC2},
from an analysis of the Miyazawa and Jernigan matrix \cite{MJ},
that the uncharged amino acids fall into two sets: hydrophobic
and polar, according to their affinity for water. Moreover,
there is experimental evidence that the native structures
of certain proteins are stable when hydrophobic amino acids are
substituted within the hydrophobic class and polar amino acids 
are substituted within the polar class\cite{Exp}.
The small number of possible interactions in an H-P model 
(H-H, H-P, and P-P) and the finite number of possible sequences
of a given length provide realistic constraints on the design 
of particular structures.
Li {\it et al.} \cite{NEC} took advantage of these constraints to 
study some design properties of an H-P model in the complete 
space of possible sequences.
In particular these authors introduced the concept of
{\it designability}.
In the terminology of Li {\it et al.},
the designability of a given compact structure
is defined as the number $N_S$ of sequences that have
this structure as their nondegenerate compact ground state;
a highly designable structure is the
nondegenerate ground state of an atypically large number of sequences.
These authors reached the conclusion that
(i) highly
designable structures are likely to be thermodynamically stable
since they have a large gap in their compact
state spectrum, 
(ii) highly designable structures are likely
to be stable against point mutations, and 
(iii) highly designable structures
have protein-like motifs.
These observations suggest that Nature's selection of protein structures
is not accidental but a consequence of thermodynamic
stability and stability against mutations.

The aim of the present article is to push further the analysis
of Li {\it et al.}. In particular, the study in \cite{NEC} is
limited to the compact-state spectra of H-P heteropolymers
of 27 monomers length
(the compact states of which fill a 3$\times$3$\times$3 cube),
and the dynamical
aspects are not discussed. Here we address 
thermodynamic and dynamic properties including all states, not only
the compact ones.
Of course, our present study is therefore limited to a small
number of sequences (45), in contrast to the complete
enumeration in \cite{NEC}.
More precisely, we address the following questions:

\begin{itemize}
\item (i) In the compact-state-spectrum studies, thermodynamic
stability is measured by the gap in the compact-state spectrum.
By gap, we mean the energy difference between the first
excited compact state(s) and the ground compact state.
How does the compact-state-spectrum gap correlate with the ``true''
thermodynamic stability where all
the possible conformations
(including the open and partially
open ones) are taken
into account? Also, how does the ``true'' thermodynamic
stability correlate with the degree of designability?

\item (ii) How does the folding time correlate with the
compact-state spectrum?
An answer to this question was postulated
in \cite{Shakhnovich1}: a small gap in the compact-state spectrum
leads to low foldability and a large gap in the compact-state spectrum
leads to fast folders. It was shown in \cite{Klimov} 
that this postulate is {\it wrong} in the
framework of a random interaction model: even though
sequences with a large compact state gap are fast folders,
sequences with a small compact state gap can fold either
slow or fast. Is the postulate right or wrong for H-P heteropolymers?

\item (iii) Are highly designable structures also fast folders ?
Sequences that have highly designable ground states
and thus are stable against mutations are also typically
thermodynamically stable according to the analysis of the
compact-state spectrum carried out in Ref. \cite{NEC}.
Moreover, Vendruscolo {\it et al.} \cite{SISSA} showed,
in different lattice models, that both types of stability are
equivalent. 
One would like to know whether the requirement of fast folding
dynamics introduces new constraints on
the set of ``protein-like" sequences. Finally, one can ask why are some sequences
fast folders while others are slow?
\end{itemize}
This article is organized as follows: section \ref{Model}
is devoted to the details of the model and the
technical aspects of the simulations. We implement
several techniques that were developed in the context of
statistical mechanics of spin models. 
These techniques are next used in sections \ref{Thermo}
and \ref{Folding} to analyze the thermodynamics
and dynamics of a set of sequences.
Section \ref{Conclusion}
summarizes our answers to the questions presented above.

\section{Model and simulation techniques}
\label{Model}
The aim of this section is to briefly recall the definition of the
heteropolymer H-P model,
and to describe the simulation techniques.

\subsection{The model}
\begin{figure}
\centerline{\psfig{file=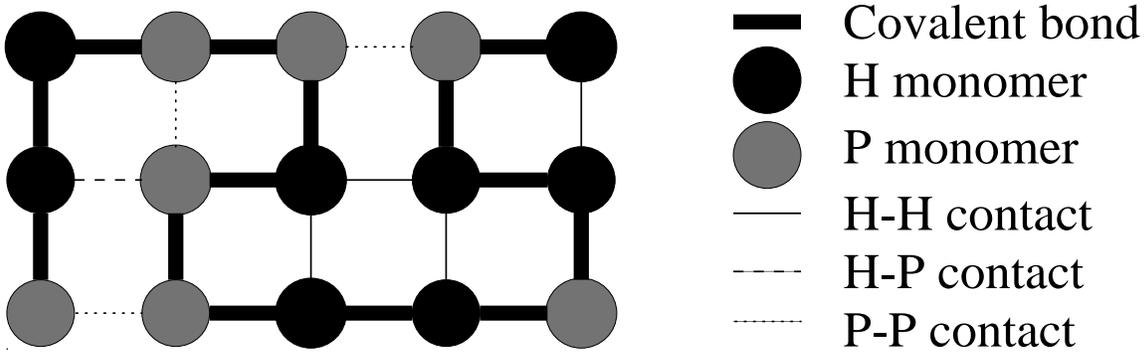,width=15cm}}
\caption{A conformation of a $15$-mer in two dimensions.}
\label{Fig0}
\end{figure}

In the model analyzed in \cite{NEC}, a protein is a
self-avoiding walk of
monomers on a cubic lattice. A sequence is defined by the set
$\{\sigma_i\}$ of amino acids along the chain, where
$\sigma_i \in \{H\mbox{(ydrophobic)},P\mbox{(olar)} \}$,
and $i$ runs over the monomers
along the chain ($i=1,...,27$ for a chain which folds into a
compact 3$\times$3$\times$3 cube).
The different
structures are defined by assigning a position ${\bf r}_i$
to the $i$-th monomer on the cubic lattice, such that
the distance between two consecutive monomers is equal to
unity and two monomers cannot lie at the same site
(self-avoidance condition). Given a sequence $\sigma$,
the Hamiltonian is
\begin{equation}
\label{Hamiltonien}
H = \sum_{i<j} E_{\sigma_i,\sigma_j} \Delta({\bf r}_i,
{\bf r}_j)
,
\end{equation}
where $E_{H,H}$, $E_{H,P}$, and $E_{P,P}$ are the energies
of H--H, H--P, and P--P contacts, respectively,
and $\Delta({\bf r}_i,{\bf r}_j)=1$
if ${\bf r}_i$ and ${\bf r}_j$ are nearest-neighboring sites
with $i$ and $j$  not adjacent along the chain,
and zero otherwise.
For instance, in the case of the two dimensional conformation
of Fig. \ref{Fig0}, there are 4 H--H contacts, 1 H--P contact,
and 3 P--P contacts and the energy
of this conformation is thus
$4 E_{H,H} + E_{H,P} + 3 E_{P,P}$.
The present model differs from the H-P models in \cite{Chan,Socci}
since
we take the interaction energies $E_{H,H}$, $E_{H,P}$, and $E_{P,P}$ to
be all different from each other.
>From the analysis of the Miyazawa-Jernigan matrix
for real proteins, it was shown in \cite{NEC2} that $E_{H,H}<
E_{H,P}<E_{P,P}$ and $(E_{H,H}+E_{P,P})/2<E_{H,P}$. This last
condition expresses the fact that different types of monomers
tend to segregate. Following \cite{NEC}, we choose
$E_{H,H}=-2.3-E_C$, $E_{H,P}=-1-E_C$ and $E_{P,P}
= -E_C$. These dimensionless coefficients define the
energy scale in which the temperature will be measured.
An increase of the compactness energy $E_C$
tends to favor compact conformations with respect
to open ones. The results of \cite{NEC} concerning the compact
state spectrum are of course independent of $E_C$.
In the present work,
the open, partially open, and compact
states are all taken into account in
the numerical calculations, so that the results presented here
depend on $E_C$.
The determination of $E_C$ from an analysis
of the Miyazawa and Jerningan matrix along the lines of \cite{NEC2}
is rather imprecise, but such an analysis suggests that $E_C$
is of the order of $2$ or $3$.
>From the point of view of our calculations, we have chosen
the overall drive to compactification $E_C$ to be large enough
so that (i) the average compactification time (being the average
time to first reach a compact state)
is much smaller
than the average folding time (being the average time
to first reach the ground state) and
(ii) very few sequences have a non-compact ground state.
However, $E_C$
should not be too large since, as was shown in \cite{Socci},
as $E_C$ increases above a certain threshold, the folded phase
gives way to a collapsed glassy phase. In practice,
after preliminary simulations, we have chosen $E_C=1.5$.

\subsection{Simulation techniques}
\subsubsection{Dynamics}
\label{BKL-sec}
The most commonly used algorithm in 
three-dimensional lattice-heteropolymer folding dynamics is the
Monte-Carlo algorithm with two one-monomer moves (end moves,
corner moves) and one two-monomer move (crankshaft move)
[see \cite{Socci} for a pedagogical description of this algorithm].
However, for the model we simulated,
this algorithm spends most of the time refusing moves, leading
to large computation times.
Hence, we chose to implement a Bortz-Kalos-Lebowitz (BKL) 
type algorithm \cite{BKL}, an algorithm especially efficient
in the presence of slow relaxation.
In contrast to the standard
Monte-Carlo algorithm, this algorithm never rejects
moves. In practice, one keeps track of all the possible states $\alpha$
that can be reached after the three types of moves listed above,
starting from a state $\alpha_0$.
Once the
list of all possible moves is established, together with the
Monte-Carlo transition probabilities,
\begin{equation}
P_{\alpha_0 \rightarrow \alpha} = \mbox{Min} \left(
\exp{\left(- \frac{\Delta E_{\alpha_0 \rightarrow \alpha}}{T}
\right) } , 1 \right)
,
\end{equation}
one move is picked
according to its relative transition probability
and that move is then performed.
The time spent in the state $\alpha_0$
is randomly chosen from the exponential distribution
\begin{equation}
P(\tau_{\alpha_0}) = \exp{\left( - \left( \sum_{\alpha \ne \alpha_0}
P_{\alpha_0 \rightarrow \alpha} \right) \tau_{\alpha_0} \right)}
.
\end{equation}
One need not recalculate this list of
possible final states at each step,
but instead one updates this list by canceling moves
that are no longer
possible and adding new moves that become possible. The energy
costs $\Delta E_{\alpha_0 \rightarrow \alpha}$
of all these moves are also updated.
This algorithm is
especially efficient for slow and average folders, but still
does not allow systematic studies at low temperatures. One
unit time of this BKL algorithm (noted BKL unit in the following)
corresponds to 27 Monte-Carlo-Steps
(MCSs) of the usual Monte-Carlo algorithm
(in lattice heteropolymer folding and for the usual Monte-Carlo
algorithm, one MCS usually corresponds to attempting
to move one monomer).

\subsubsection{Thermodynamics}
\label{histo}
One possibility to compute thermodynamic quantities is to carry
out an exhaustive enumeration of all the possible conformations,
including the noncompact ones. Such an exact method
was used for instance in \cite{Klimov} in the case of a 15-mer
in three dimensions, and in \cite{SISSA} for a 16-mer
in two dimensions. Since, in practice, we could not apply
this exact method to 27-mers in three dimensions, we have used
Monte-Carlo histogram techniques to
study the thermodynamics.

Several groups working on proteins
have rediscovered the Monte-Carlo histogram technique and applied it
to heteropolymer models of proteins.
We refer the reader to \cite{Socci} for a clear description of
this technique. Most simply, the technique consists of
carrying out a simulation at a given temperature $T_0$ and keeping track
of the histogram of various quantities, for instance the joint
probability of the energy $E$, number of contacts,
and similarity with the
ground state. Using this histogram calculated at $T_0$,
one can recalculate several thermodynamic quantities at another
temperature $T$
by changing the Boltzmann weight to $\exp{(-E/T)}$
without carrying out a new simulation.
However,  $T$ should remain close to $T_0$ since the
simulation at temperature $T_0$ is carried
out over a finite time and
the phase space is thus only partially sampled.
We did not in fact use this ``naive'' histogram technique
but rather made use of the powerful method invented
by Ferrenberg and Swendsen \cite{FS,Mul-His}.
This method consists in optimizing
several histograms calculated at different temperatures
to obtain the temperature dependence of various thermodynamic
quantities. Moreover, the accuracy of the optimized
thermodynamic quantities can be evaluated, and additional
simulations can be carried out when necessary.
Following \cite{FS},
we briefly summarize this technique by focusing
on the case of the single energy histogram,
the generalization to joint histograms of various
quantities being straightforward.
We consider $R$ energy histograms labeled by $n=1,..R$,
carried out over $t_n$ time units 
at the temperatures $T_1,...,T_n,...,T_R$, with the
normalization
$$
t_n = \sum_E N_n(E)
,
$$
where $N_n(E)$ is the number of times a state (or states) with energy
$E$ is sampled in the $n$-th histogram.
The partition function $Z(T)$ is expanded over the density
of states $W(E)$ as
$$
Z(T) = \sum_E W(E) \exp{(-E/T)}
.
$$
Each of the $R$ histogram simulations carried out at a
temperature $T_n$ leads to a ``naive'' estimate
of the density of states
\begin{equation}
\label{naive-W}
W_n(E) = t_n^{-1} N_n(E) \exp{( E/T_n - F_n/T_n)}
,
\end{equation}
with $F_n$ the free energy at temperature $T_n$.
The Ferrenberg and Swendsen method \cite{FS} consists
in expanding the density of states $W(E)$ as a combination
of the ``naive'' densities of states (\ref{naive-W})
$$
W(E) = \sum_{n=1}^R p_n(E) W_n(E)
,
$$
the coefficients of this expansion (as well as
the free energies $F_n$ in (\ref{naive-W})) being determined 
by minimizing the error in this estimation of $W(E)$.
The result is a closed set of multiple-histogram equations for 
the $F_n$ \cite{FS}
\begin{eqnarray}
\label{self1}
P(E,T) &=& \frac{
\sum_{n=1}^R N_n(E) \exp{(-E/T)} }
{ \sum_{n=1}^R t_n \exp{(-E/T_n + F_n/T_n)} }\\
\exp{(F_n/T_n)} &=& \sum_E P(E,T_n)
.
\label{self2}
\end{eqnarray}
Once (\ref{self1}) and (\ref{self2}) have been
solved by successive iterations, the average of
an energy-dependent operator is calculated as
$$
\langle {\cal O}(E) \rangle =
\frac{\sum_E {\cal O}(E) P(E,T)}
{\sum_E P(E,T)}
.
$$

\section{Thermodynamics}
\label{Thermo}
The compact state spectrum was previously determined by means of
exact enumerations \cite{NEC}. Thanks to this work, we know
for each sequence the lowest energy compact state.
However, it is possible that the true ground state is not compact.
We therefore checked during the histogram calculations that no
open state has a lower energy than the compact ground state.
We eliminated a few sequences with noncompact ground states.
Since we were not able to perform an exact enumeration of the
noncompact states, we cannot exclude
the possibility of a noncompact ground state that
was not found during the simulations. However, we believe that this
possibility is rather unlikely.
We will comment later on the characteristics of those few sequences
with a noncompact ground state.

\subsection{Qualitative study of a thermodynamically stable
and unstable sequence}
\label{qualitative-study}
We first begin with a qualitative analysis of a ``protein-like''
and a ``non-protein-like'' sequence, from the point of view of
thermodynamics. As stated in the introduction,
one necessary condition for a sequence
to be ``protein-like'' is that it be thermodynamically stable. Hence,
we have to find a quantitative way to measure thermal stability.
It has been proposed \cite{Goldstein} that real proteins undergo
a folding transition at some temperature $T_f$ larger than the
glass transition $T_g$, below which the dynamics dramatically freezes.
For ``protein-like'' sequences,
folding should be a first-order-like transition,
with $T_f$ much larger than
$T_g$, thus allowing the possibility of a temperature regime in which
the ground state is thermodynamically favored and kinetically
accessible.
By ``first-order-like'', we mean that as a function of temperature the
native state occupancy has a narrow transition from a
low value in the unfolded phase to a high value
in the folded phase. We will use the
width of this transition as a  measure of
thermodynamic stablility, the
thermodynamically stable sequences having a narrow
transition.
The first-order-like behavior of the folding
transition was put on a quantitative basis
for a lattice H-P model by Socci and Onuchic \cite{Socci}.
These authors studied the shape of the energy histogram and observed
a transition from a unimodal shape to a bimodal shape
as the temperature was lowered below $T_f$, suggesting the
occurrence of a first-order-like transition.

On the other hand, in the case of a thermodynamically unstable sequence
(and thus one that is not ``protein-like''), there are low-lying
states competing with the ground state, and the transition to
the folded ground state as the temperature is decreased is thus broad.
In order to compare different sequences,
we measure thermodynamic stability in terms of the width
in temperature
of the transition to the native state compared to the
transition temperature.
In biological situations, the temperature is usually fixed
and the notion of thermal stability is then not only sequence-dependent
but also temperature-dependent.
For instance, some sequences could have a narrow transition
to the folded state but with  too low a $T_f$,
so that in practice, the protein is unfolded at the
temperature of interest. Our definition of
thermodynamic stability is thus, in the context of real proteins,
only a necessary condition. 

\begin{figure}
\centerline{\psfig{file=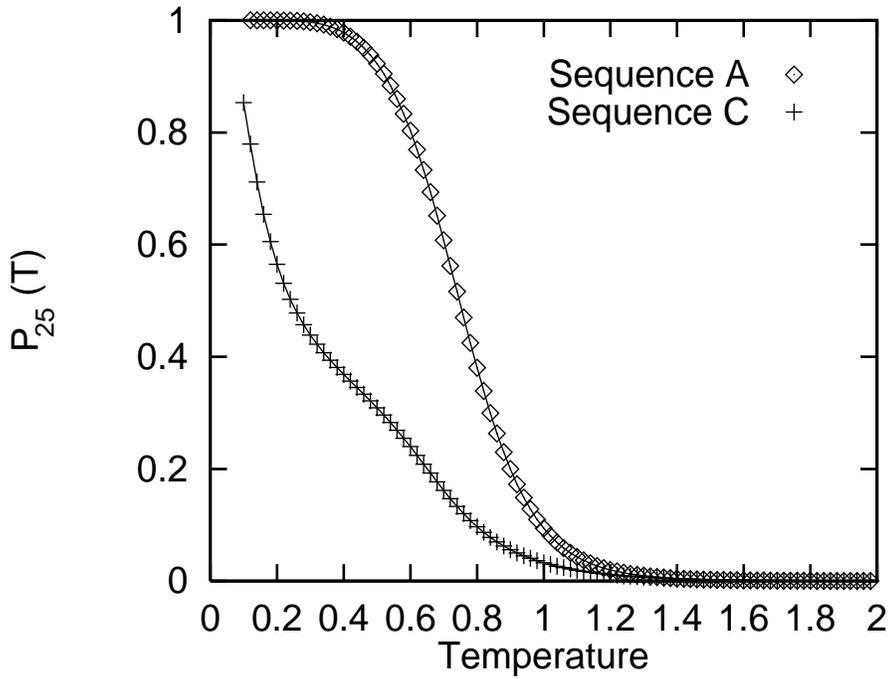,height=8cm}}
\caption{Probability ${\cal P}_{25}(T)$ to be in any state having
more that $25$ native contacts as a function of temperature
for the thermodynamically stable sequence A and the thermodynamically
unstable sequence C.}
\label{Fig1}
\end{figure}

Following Socci and Onuchic \cite{Socci}, we employ
the optimized histogram method described in section
\ref{histo},
and determine the probability ${\cal P}_{25}(T)$
that the sequence is in any conformation
having more than $25$ correct ground-state contacts, 
or ``native" contacts,
out of $28$ possible contacts for the fully folded structure.
Throughout this paper, this set of conformations will be referred
to as ``native states''.
We will consider a heteropolymer as correctly 
folded if the number of native contacts is larger than or equal
to $25$. This assumption speeds up computations and
is sensible from the point of
view of real proteins since structure fluctuations
around the folded conformation are expected,
and do not prevent the protein
from being functional (see for instance \cite{Petsko}).
The variation of ${\cal P}_{25}(T)$ as a function of temperature $T$
are shown on Fig. \ref{Fig1}
for the thermodynamically stable sequence A and the
thermodynamically unstable sequence C. [These two sequences
will be studied in detail in the present article].
The larger width of the transition to the folded conformation
in the case of sequence C  is due to the existence of low-lying
semi compact conformations, not present for thermodynamically
stable sequences. The consequences
of these low-lying states for the folding dynamics will
be investigated in what follows.

\subsection{Thermodynamic stability: compact state spectrum
versus transition width}
In order to answer the first question in item (i) in the
introductory section, we want to compare the
compact-state-spectrum estimation of the thermodynamic stability,
in terms of the folding temperature $T_f$ and the glass-transition 
temperature $T_g$ \cite{Goldstein,Bryngelson1},
to the thermodynamic stability estimated with the method of section
\ref{qualitative-study}.

\subsubsection{Compact-state-spectrum results}
We first review the results of Goldstein {\it et al.}
\cite{Goldstein}, and Bryngelson {\it et al.} \cite{Bryngelson1},
who estimated the folding
temperature $T_f$ and the glass temperature $T_g$ in terms
of simple spectral quantities. These authors derived their
criteria from the complete energy spectrum,
dividing phase space into two components:
on the one hand, the native
state with a nearly zero entropy and, on the other hand,
uncorrelated ``liquid-like'' states at higher energy with a
large entropy. If $\Delta$ denotes the
average energy difference between the native
and the liquid-like states, and $\Gamma$ is the width in
energy of the liquid-like states,
it is shown in \cite{Goldstein,Bryngelson1}
that the ratio $T_f/T_g$ is maximal if $\Delta/\Gamma$ is maximal.
In other words, the ratio $\Delta/\Gamma$ should measure
how ``protein-like'' a sequence is.
We take
here as a working hypothesis the application of
the results in \cite{Goldstein,Bryngelson1} to 
our lattice model including the compact states only.
In order to test the hypothesis,
we calculate $\Delta$ and $\Gamma$ from
the compact-state spectra, the compact states being
conformations exactly filling the 3$\times$3$\times$3 cube,
and the compact-state spectrum of a given sequence being
the set of energies of all compact conformations.
More precisely,
\begin{eqnarray}
\label{Delta}
\Delta &=& \frac{1}{N_C} \sum_{\alpha>0} (E_{\alpha} - E_0)\\
\Gamma^2 &=& \frac{1}{N_C} \sum_{\alpha>0} E_{\alpha}^2
- \left( \frac{1}{N_C} \sum_{\alpha>0} E_{\alpha} \right)^2
\label{Gamma}
,
\end{eqnarray}
where the $E_{\alpha}$ ($\alpha>0$) denote the energies of the
excited compact conformations, $E_0$ is the lowest compact
state energy, and $N_C$ is the number of excited compact conformations.

\subsubsection{Thermodynamic stability versus
$\Delta/\Gamma$}
\begin{figure}
\centerline{\psfig{file=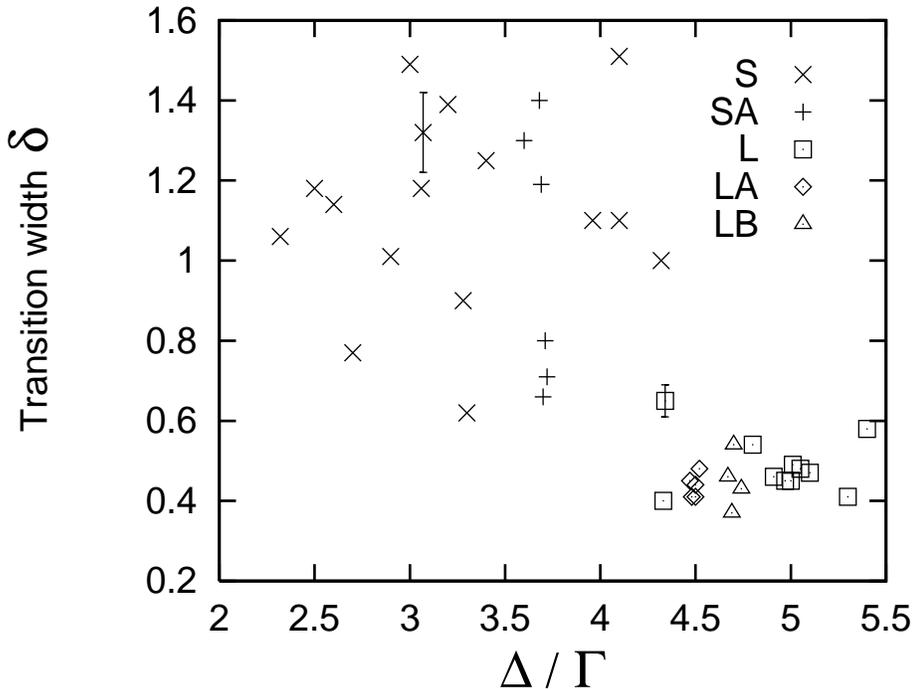,height=8cm}}
\caption{Transition width $\delta$ from the
Monte-Carlo simulations versus $\Delta/\Gamma$ obtained   
from the compact-state spectrum. As explained in the text, the
different sets and sub-sets of sequences are denoted by
different symbols. Typical error bars in the estimation
of $\delta$ are shown. The larger error bars for large
$\delta$ originate from the increase of glassiness
as $\delta$ increases.}
\label{Fig4}
\end{figure}

As far as the Monte Carlo results are concerned, we estimate
the thermal stability of a sequence
by the width of the transition
in ${\cal P}_{25}(T)$ {\it versus} $T$, as shown
in Fig. \ref{Fig1}. More precisely, we calculate
the dimensionless transition width as
\begin{equation}
\label{delta-def}
\delta=2 \left( \frac{T(0.1)-T(0.8)}{T(0.1)+T(0.8)} \right)
,
\end{equation}
where $T(x)$ is the temperature for which the
probability is $x$ to
find the heteropolymer in a state with $25$ or more native
contacts. [This is the inverse function of
${\cal P}_{25}(T)$ as plotted on Fig. \ref{Fig1}].
We have plotted in Fig. \ref{Fig4} the width $\delta$ of the
transition as a function of
$\Delta/\Gamma$ defined by (\ref{Delta}) and (\ref{Gamma}).
We have separated the sequences into a set
``L(arge)''of sequences with
large values of $\Delta/\Gamma$  and
a set ``S(mall)'' with small values of this ratio.
(For further consideration, we have extracted from the ``L'' set
two subsets ``LA'' and ``LB'' with a fixed 
$\Delta/\Gamma$ ratio, of order $4.5$ and $4.7$
respectively. Also, the set ``SA''  has been extracted from
the ``S'' set, with a $\Delta/\Gamma$ ratio of order $3.7$.)
It is clear in Fig. \ref{Fig4} that
in spite of some scatter, the ``L'',
``LA'' and ``LB'' sets, with a high
$\Delta/\Gamma$ ratio (larger that $4.3$) correspond
to thermodynamically stable sequences,
with a narrow transition to the
native structure ({\it i.e.}, a small $\delta$ value),
whereas the transition to the native structure
for the sequences belonging to the ``S'' and ``SA'' sets
($\Delta/\Gamma < 4.3$) is rather broad. This shows
that ``protein-like'' sequences (at least as far as thermodynamics
is concerned) correspond to sequences with a large $\Delta/\Gamma$
ratio, {\it i. e. } sequences with a low-energy native state
compared to the energy width of the distribution of compact
states.

\begin{figure}
\centerline{\psfig{file=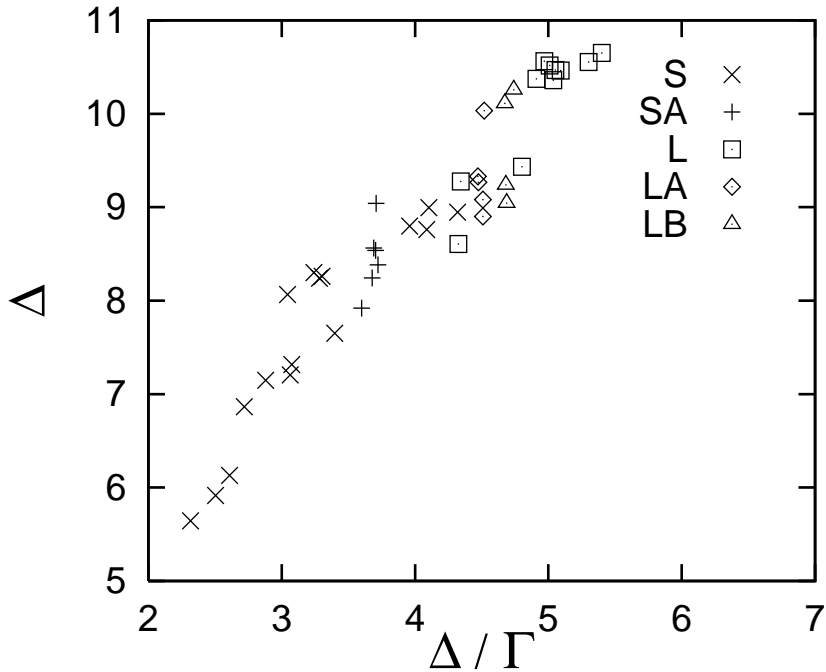,height=8cm}}
\caption{$\Delta$ {\it versus} $\Delta/\Gamma$ for
our 45 sequences selection. The symbols are the same
as in Fig. \protect \ref{Fig4}.
}
\label{Fig-Delta}
\end{figure}

One can also ask how the transition width $\delta$
correlates with $\Delta$ alone, independent of
$\Gamma$. It turns out that
$\Delta/\Gamma$ is almost a monotonically increasing function
of $\Delta$, as shown in Fig. \ref{Fig-Delta},
$\Gamma$ being nearly constant at $\Gamma \simeq 2$,
and therefore $\Delta$ alone is also a good predictor of
thermodynamic stability.

\subsection{Thermodynamic stability versus designability}
The designability of a given structure is measured in terms of
the number $N_S$ of sequences that have this structure as
their nondegenerate ground state, and, as proposed in
\cite{NEC}, the sequences corresponding to 
highly designable structures should be
``protein-like''.
Interestingly, the highly designable structures show regular helix-like and
$\beta$-sheet-like patterns \cite{NEC}, quite reminiscent
of the regular patterns in real proteins.
Moreover, since these highly designable structures have on
average a large gap to their first compact excited state
\cite{NEC}, they are likely to be thermodynamically stable.
More specifically, the average gap in the
compact state spectrum suddenly jumps from a low value to a high value
as $N_S$ increases above $N_S \sim 1400$.
It is thus of interest to relate the thermodynamic stability
of sequences, measured in terms of $\delta$ in equation
(\ref{delta-def}), to the degree of designability of their
native state structure measured by $N_S$. We will
reach the conclusion that $\Delta/\Gamma$ is a better
predictor of thermodynamic stability than designability.
Qualitatively, this could be anticipated since
high designability of a structure still
allows for very different behaviors of the sequences which
design that structure.

\begin{figure}
\centerline{\psfig{file=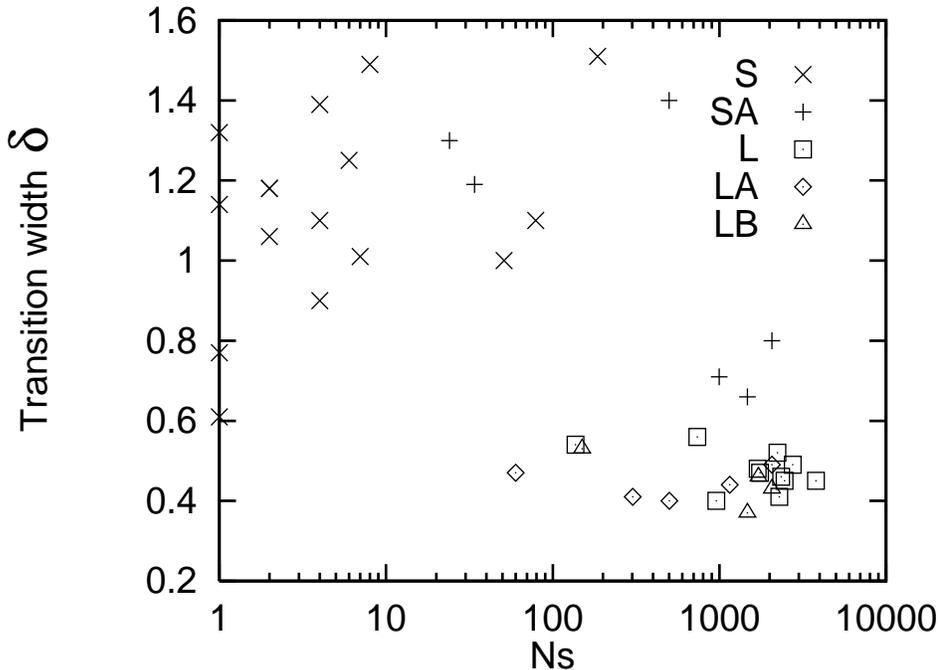,height=8cm}}
\caption{Transition width $\delta$ versus the number $N_S$ of
sequences that design a given structure. Crosses denote
sequences with a small $\Delta/\Gamma$ ratio
and squares, diamonds and triangles sequences with
a large $\Delta/\Gamma$. Some sequences in the SA
sub-set with $\Delta/\Gamma \simeq 3.7$ can have
a large designability in spite of being thermodynamically
unstable.
}
\label{Fig5}
\end{figure}

\begin{figure}
\centerline{\psfig{file=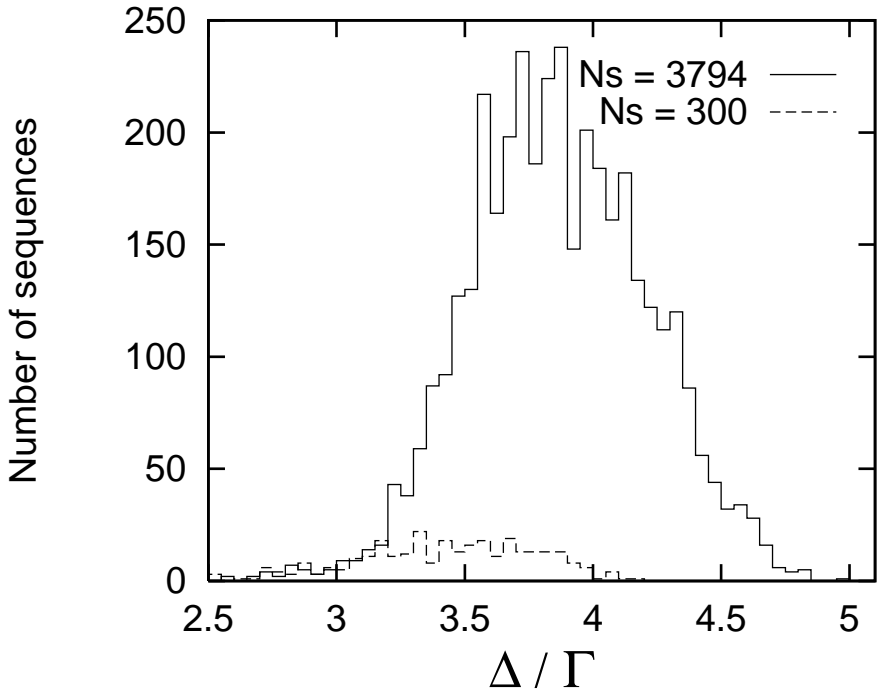,height=8cm}}
\caption{Distribution of $\Delta/\Gamma$ for the top structure
(structure with the maximal $N_S=3794$, solid line)
and a structure with $N_S=300$ (dashed line).
Each bin has a width of $0.05$ in $\Delta/\Gamma$ and the
quantity on the $y$ axis measures how many sequences fall in
each bin.
}
\label{Fig6}
\end{figure}

The correlation between the transition width $\delta$ of
the various sequences analyzed here,
and the number $N_S$ of sequences that design their
native structures is plotted in Fig. \ref{Fig5}.
Clearly, the sequences whose ground states are
highly designable structures
are likely to be thermodynamically stable.
However, we observe that several sequences belonging
to the sets ``S'' and ``SA'' of thermodynamically unstable
sequences have a relatively high designability $N_s$.
These sequences, despite high designability,
have a small value of $\Delta/\Gamma$.
Such sequences are expected
to emerge due to the fact that, even for a structure
with large $N_S$, the ratio
$\Delta/\Gamma$ is not necessarily large
for all sequences having that structure as a ground state \cite{voron}.
In other words, it is possible to find atypical sequences
with a low value of $\Delta/\Gamma$ even for structures
with a large $N_S$.
For instance, using the compact-state spectra,
we have plotted in Fig. \ref{Fig6} the distribution
of $\Delta/\Gamma$ for the top structure (this is the
structure with the maximal $N_S=3794$)
and a structure with $N_S=300$.
On average the sequences corresponding to the structure with 
$N_S=3794$ have larger $\Delta/\Gamma$ than the sequences  
corresponding to the structure with $N_S=300$. However,  
the tail of the distribution for the $N_S=3794$ structure  
still extends to very small $\Delta/\Gamma$.
In summary, large $\Delta/\Gamma$ ratio is a good predictor that
a sequence will be thermodynamically stable. A high 
designability $N_S$ for a structure is consistent with, but
does not guarantee, large $\Delta/\Gamma$ and hence thermodynamic
stability of its associated sequences.

Finally, we would like to comment on the results of Vendruscolo
{\it et al.} \cite{SISSA}. These authors also carried out
an analysis of the relation between thermodynamic stability,
calculated with all the configurations including the open ones,
and the number of sequences $N_S$ that design a given structure.
The authors analyzed the two dimensional case of the present model
with 16 monomer chains and exactly
enumerated not only all the possible
compact but also all the possible open state conformations.
They found first, in agreement with Li {\it et al.} \cite{NEC},
a broad distribution of $N_S$. However,
they showed that the thermodynamic stability is flat
as a function of $N_S$, in contradiction with our present
results, and also in contradiction with unpublished
data \cite{NEC-unpub} where the gap in the compact state spectrum
in two dimensions was shown to
increase as a function of $N_S$.
In our opinion, this discrepancy is due to
the fact that in \cite{SISSA} the overall compactness energy $E_C$
in the inter-residue interactions 
was set to zero, thereby weighting open state
configurations too strongly with respect to compact states.
We expect that a larger $E_C$
would have led to an $N_S$-dependence of
the thermodynamic stability in Ref. \cite{SISSA}.

\subsection{Conclusions on the thermodynamics}
This section was devoted to the analysis of the thermodynamics of
several sequences, selected from highly and poorly
designable sequences, and with large and small values
of $\Delta/\Gamma$. The known compact state spectra allow
for the calculation of the ratio $\Delta/\Gamma$,
which is thought to be proportional
to the ratio of the folding temperature to the glass temperature
\cite{Goldstein,Bryngelson1},
and thus predicts to what extent a given sequence is
thermodynamically ``protein-like''.
The Monte-Carlo data calculated
with open and partially open states also allowed us to characterize the
thermodynamic stability of a given sequence, in terms of the width
$\delta$ of its folding transition. We have shown that these two quantities,
$\delta$ and $\Delta/\Gamma$, 
are well correlated, signaling the validity of the compact-state-spectrum 
analysis. We have also shown that stability
with respect to mutations (measured in terms of
designability) is not directly equivalent to thermal stability.

Finally, it is worth noting that we eliminated three
sequences with partially open
ground states. These sequences had a very low
$\Delta/\Gamma$ ratio ($2.17$, $2.18$, $2.52$) [this ratio has been
calculated with the ground state energy $E_0$ being the lowest
compact state energy]. By contrast, none of the simulated
sequences with a high or moderate
$\Delta/\Gamma$ ratio were found to have
a partially open ground state. This suggests that
the ground state conformation of
some of the remaining sequences
with a small $\Delta/\Gamma$ ratio may not be compact. However,
since these sequences are among the most unstable ones, our calculation
of the transition widths as far as ``protein-like''
sequences are concerned is unaffected.

\section{Folding simulations}
\label{Folding}
Besides thermodynamic stability and stability with respect
to mutations of the amino-acid code, the native state
of a ``protein-like'' sequence should be dynamically accessible.
Clearly, the dynamics of all the sequences, including the
thermodynamically stable ones, becomes glassy at low temperatures
in our model
since all the inter-residue interactions are attractive.
As the temperature
is lowered, the average
folding time (the average time required to first reach the native
states, starting from a stretched conformation) will first
decrease and then, below a certain temperature, will start to
increase drastically, presumably scaling like
$\ln{(\langle \tau_f \rangle)} \sim 1/T$. Conversely, as
we have already seen, the native state occupancy increases
as the temperature is decreased. Since
we want to investigate the balance between thermodynamic
stability and dynamical accessibility of the native state,
we chose to examine the dynamics at the
temperature $T_f$ such that the
native state occupancy is fixed at 
some predetermined ${\cal P}_f$.
The question
is then: do the sequences with a large $\Delta/\Gamma$
ratio, thermodymic stability, and mutational stability
also fold fast at $T_f$ ?

In order to examine this question,
we measured the average folding
time, namely the average time $\langle \tau_f \rangle$
necessary for the
heteropolymer to first reach
any of the native states (with $25$ ground state contacts,
out of $28$), starting from a stretched conformation.
Because in our model all interaction energies are attractive,
and individual non-native contacts may be as strong as native
contacts,
the folding dynamics is very slow compared to
other models.
For this reason, we simulated folding of the heteropolymers
at the temperature $T_f$ such that the average occupancy of
the native states is only ${\cal P}_f = 10 \%$.
Compared to other works where
models with a faster folding time were
investigated,
this occupancy of the native states
is quite small. For instance, Abkevich {\it et al.}
\cite{Abkevich} could fold a few sequences at a
temperature for which the average similarity with ground state
was up to $95 \%$. Their slowest average folding time was then
of the order of $2 \times 10^8$ MCS, which is one order of
magnitude smaller than the fastest folding time
in our simulations (given that
$1 \mbox{ BKL unit}= 27 \mbox{ MCS}$). 
However, models with faster folding times generally require unrealistic
assumptions such as artificially lower energy for individual
native contacts compared to the non-native ones. 
Even though we
could not go systematically to lower temperatures,
the effect of lowering the temperature was examined
for a few sequences.

\subsection{Average folding times}
\begin{figure}
\centerline{\psfig{file=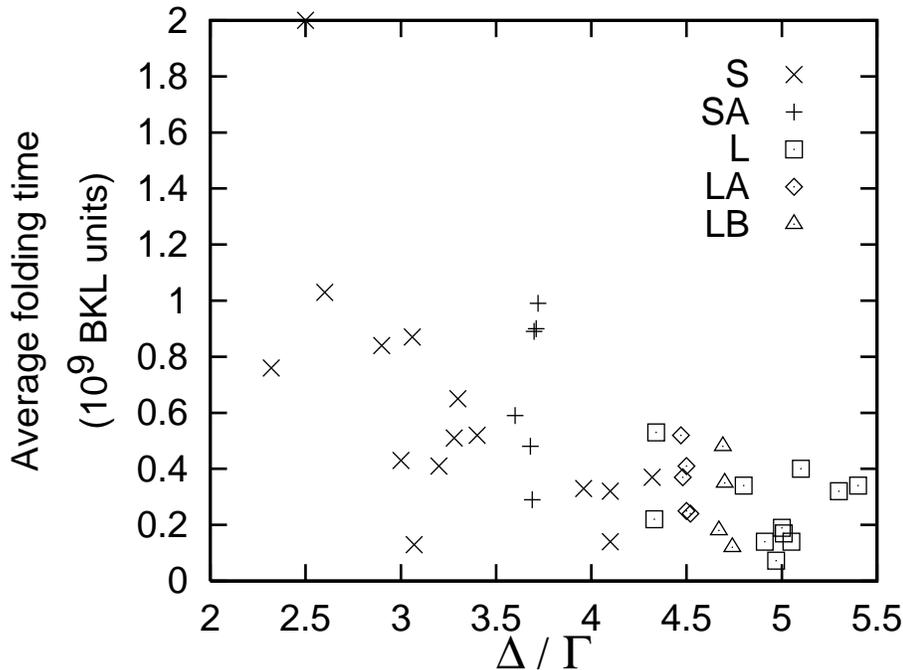,height=8cm}}
\caption{Average folding times (in BKL units) versus $\Delta/\Gamma$,
 with the same symbols as Figs.
\protect\ref{Fig4}, \protect\ref{Fig-Delta},
and \protect\ref{Fig5}.
A very slow folding sequence
(hereafter denoted by ``W(orst)'') has
not been shown. It belongs to the set ``S'',
has $\Delta/\Gamma = 2.7$, and an average
folding time of $4.3 \times 10^9$ BKL units.
}
\label{Fig7}
\end{figure}

One folding simulation consists in starting from a stretched
heteropolymer, and running the dynamics until one of the native
states is first reached. The time required is the
folding time, which should then be
averaged over several folding simulations. In practice, we
have averaged the folding time over 20 folding simulations.
For each sequence, the temperature is fixed at $T_f$ (at this
temperature, the
native-states occupancy is ${\cal P}_f = 10 \%$).
The average folding times are plotted in Fig. \ref{Fig7} as a
function of $\Delta/\Gamma$.
The relative error in the average folding time
can be estimated by noticing that the mean value of
the folding time distribution is of the order of
the standard deviation (we indeed calculated
the full folding time distribution for a
few sequences). The relative error in the estimation
of the average folding time is then of the
order of the inverse square root of the number of
folding simulations, namely, in our case of
$1/\sqrt{20} \simeq 0.2$. This precision is
sufficient for the purpose of our discussion.

We observe in Fig. \ref{Fig7} that sequences
with a large $\Delta/\Gamma$ ratio fold fast.
We thus conclude, in response to question (iii) in the
introductory section, that folding dynamics does not add any
constraint in the selection of
``protein-like'' sequences: once a structure
is stable against mutations and thermodynamically stable
(namely, a sequence with a large $\Delta/\Gamma$ ratio
and a large $N_S$), it will
be a fast folder.

A quite striking result visible in Fig. \ref{Fig7}, is that
sequences with a low $\Delta/\Gamma$ ratio (belonging to 
the ``S'' and ``SA'' sets of thermodynamically unstable sequences),
may also fold fast, at least as fast as some of the sequences
belonging to the ``L'', ``LA'' or ``LB'' sets.
However,
at lower temperatures and hence larger native-states
occupancy, we find a stronger correlation between
low $\Delta/\Gamma$ ratio and slow folding.

In order to go to lower temperatures, we
selected four sequences (denoted by A, B, C, and D;
A and C being the same sequences as in Fig. \ref{Fig1}),
with
the same fast folding time, $0.14 \times 10^9$ MCS,
at $T_f$, i.e. with a $10\%$ occupancy of conformations with
25 or more native contacts. We
folded these sequences again at temperatures corresponding to
larger native-states occupancy.
\begin{figure}
\centerline{\psfig{file=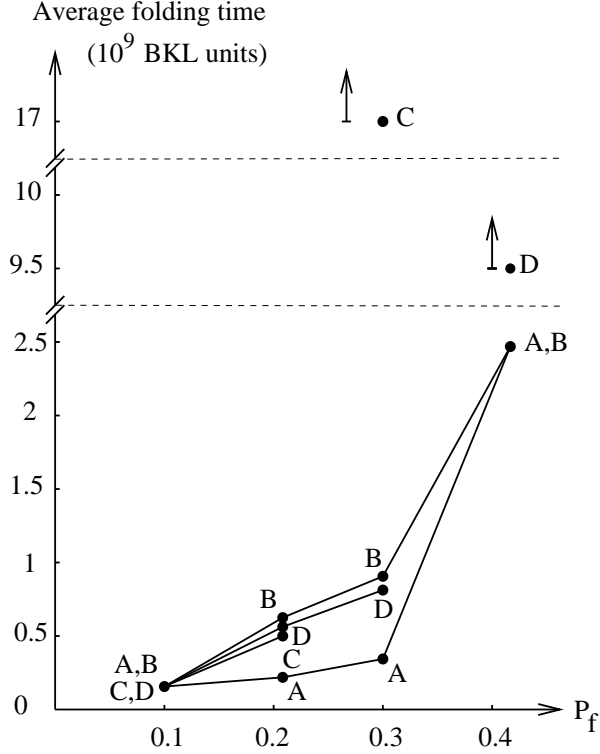,height=10cm}}
\caption{Folding time of the four sequences A, B, C and D
for various native-states occupancies.
The value of $\Delta/\Gamma$ for the four sequences
A, B, C and D is respectively $5.05$, $4.91$,
$4.1$ and $3.07$. The sequences A and B are thermodynamically
stable, and C and D are thermodynamically unstable.
The two arrows indicate that the corresponding folding
times are a lower bound.
}
\label{Figsupp}
\end{figure}
The sequences A and B belong to the set ``L'' of thermodynamically
stable sequences, and the sequences C and D belong to the set ``S''
of thermodynamically unstable sequences.
As the native-states occupancy is increased,
we observe in Fig. \ref{Figsupp} that the folding times
of the two thermodynamically unstable sequences, C and D,
clearly become much larger than the
folding times of the two thermodynamically
stable sequences, A and B. Regarding the results of
our study at a fixed
native-states occupancy of $10 \%$ (see Fig. \ref{Fig7}),
this suggests that, as the native-states occupancy increases,
the average folding times of the sequences with small
$\Delta/\Gamma$ ratio will increase much more rapidly
than the folding time of the sequences with
large $\Delta/\Gamma$ ratio.

The conclusion regarding
the variations of the average folding times as a function
of the native-states occupancy is similar to the
one reached by Abkevich {\it et al.} \cite{Abkevich}, who
proposed that the gap $\Delta$ in the compact-state spectrum
correlates with the average folding time
calculated for large native-states occupancy.

\subsection{Slow versus fast folding}
Despite the complexity of protein energy landscapes,
several authors have proposed simple effective
descriptions of these landscapes. For instance,
the role of traps (or folding intermediates) in the folding 
process was underlined by several authors, for
instance by Bryngelson and Wolynes \cite{Trap} and
by Klimov and Thirumalai \cite{Klimov}.
A scenario was proposed by Bryngelson {\it et al.}
\cite{pathway} in which there exists a ``folding pathway''
to the native state: in a first step, the protein collapses
via many possible paths in phase space, and, in a second step,
folds to the native structure via a small number of possible
paths. Abkevich {\it et al.} found evidence for a nucleation process in
a lattice model study \cite{nucleus}. The aim of the present section is
to see if there exists any simple scenario in our H-P model.

Among all the sequences analyzed here at a ${\cal P}_f = 10 \%$
occupancy of native states,
one sequence has an anomalously long folding time,
approximatively $4 \times 10^9$ MCS.
This sequence belongs
to the set of thermodynamically unstable sequences
($\Delta/\Gamma = 2.7$ and $\delta = 0.77$).
Since this sequence has the longest folding time among our
sequence selection, we will denote this sequence
by ``W(orst)''.
We will compare this very slow folding
sequence to a fast folding 
``protein-like'' sequence (sequence A in
Fig. \ref{Figsupp}),
and also to a sequence which is 
thermodynamically unstable but folds fast at a native-states
occupancy of ${\cal P}_f = 10 \%$ 
(sequence C in Fig. \ref{Figsupp}).

\subsubsection{Number of low-energy conformations}
\label{clusters}
In order to obtain more information on low-energy traps, 
we carry out the following simulation.
We start from a stretched conformation and let the dynamics
evolve until a conformation is first reached whose energy above
the ground state
$\Delta E = E - E_0$ is smaller than a given
$\Delta E_{\rm \, hit}$. 
We refer to the result of a single such simulation as a ``hit".
The contact matrix of this conformation is then
recorded. (The contact matrix encodes a compact or nearly compact structure
in a unique way. Its matrix elements
${\cal C}_{i,j}$, with $i,j = 1,...,27$, labeling the
monomers along the chain, are 
equal to unity if
the mononers $i$ and $j$ are in contact, and zero otherwise.)
We repeat this simulation $N$ times ($N = 1500$ in practice).
We have chosen a small $\Delta E_{\rm hit} = 0.5$,
but similar conclusions were
obtained with $\Delta E_{\rm hit} = 1$. 
We finally examine the different structures encoded
in the set of contact matrices
${\cal C}^{(1)}$, ... , ${\cal C}^{(N)}$ of the $N$ ``hits''.

We analyzed three sequences: sequence A
(a ``protein-like'' sequence), sequence C (fast folding at
${\cal P}_f = 10 \%$ occupation of native states,
but thermodynamically unstable),
and the very slow folding sequence W.

We first study the number of different
structures for the sequences A, C, and W. It turns
out that the ``protein-like''
sequence A,
no states other than the ground state were       
found with an energy $\Delta E < \Delta E_{\rm hit}=0.5$.
As far as the two other sequences C and W are concerned,
we found $12$ different structures for sequence C
and $125$ for sequence W.
The list of these structures obtained
is shown in Table \ref{table1} for 
sequence C and Table \ref{table2} for
sequence W.
The Hamming distance to the ground state in the second
column of these tables is
\begin{equation}
\label{Distance}
d({\cal C},{\cal C}^{(GS)}) = \sum_{i=1}^{27} \sum_{j=i+3}^{27}
| {\cal C}_{i,j} - {\cal C}_{i,j}^{(GS)} |
,
\end{equation}
with ${\cal C}$ the contact matrix of a given hit
and ${\cal C}^{(GS)}$ the contact matrix of the
ground state structure.
The contact matrices being symmetric, the summation in
(\ref{Distance}) is restricted to the matrix elements
$i \le j$. The matrix elements $j=i+1$ have been discarded
since they correspond to covalent bonds; the
matrix elements $j=i+2$ have also been discarded 
since no contact can be made between
these monomers.

\begin{table}
\begin{center}
\begin{tabular}{|c|c||c|c||}
\hline
Density  & $d_{GS}$ &
Density & $d_{GS}$ \\
of hits (\%)& &
of hits (\%)&  \\
\hline \hline
 21.0  &  28 	& 6.0 	& 38 \\ \hline
 17.7  &  32 	& 5.4 	& 40 \\ \hline
 10.9  &  30 	& 4.7 	& 38 \\ \hline
 10.7  &  38    & 3.9	& 34 \\ \hline
 8.7   &  26    & 2.1 (GS)& 0 (GS) \\ \hline
 7.9   &  34 	& 1.2	& 40 \\ \hline
\end{tabular}
\caption{The twelve structures obtained for sequence C.
The energy of these structures is $\Delta E=0.3$ (except
for the ground state that has $\Delta E=0$).
The first column is the density of hits (probability that
a given structure is hit) and the second
column is the Hamming distance to the ground state.
The ground state is indicated in the table.
}
\label{table1}
\end{center}
\end{table}
\begin{table}
\begin{center}
\begin{tabular}{|c|c||c|c||c|c||c|c|}
\hline
Density & $d_{GS}$ &
Density & $d_{GS}$ &
Density & $d_{GS}$ &
Density  & $d_{GS}$\\
of hits (\%)&  &
of hits (\%)&  &
of hits (\%)&  &
of hits (\%)&  \\
\hline \hline
$\mbox{ }$ 4.7 & 46 &$\mbox{ }$ 1.0 & 42 &$*$ 0.5 & 32 &$*$ 0.3 & 42 \\ \hline
$\mbox{ }$ 4.5 & 44 &$*$ 1.0 & 42 &$*$ 0.5 & 40 &$*$ 0.3 & 48 \\ \hline
$\mbox{ }$ 3.4 & 48 &$\mbox{ }$ 0.9 & 46 &$\mbox{ }$ 0.5 & 42 &$*$ 0.3 & 50 \\ \hline
$\mbox{ }$ 3.3 & 44 &$*$ 0.9 & 44 &$\mbox{ }$ 0.5 & 44 &$*$ 0.3 & 44 \\ \hline
$\mbox{ }$ 3.3 & 46 &$*$ 0.9 & 34 &$*$ 0.5 & 12 &$*$ 0.3 (GS)&  0 (GS) \\ \hline
$\mbox{ }$ 3.1 & 46 &$\mbox{ }$ 0.9 & 42 &$*$ 0.5 & 50 &$\mbox{ }$ 0.3 & 38 \\ \hline
$\mbox{ }$ 2.1 & 42 &$\mbox{ }$ 0.9 & 44 &$*$ 0.5 & 42 &$*$ 0.3 & 38 \\ \hline
$\mbox{ }$ 2.1 & 42 &$*$ 0.9 & 40 &$\mbox{ }$ 0.5 & 46 &$*$ 0.2 & 40 \\ \hline
$\mbox{ }$ 2.1 & 48 &$*$ 0.9 & 46 &$\mbox{ }$ 0.5 & 46 &$*$ 0.2 & 34 \\ \hline
$*$ 2.0 & 46 &$*$ 0.9 & 38 &$\mbox{ }$ 0.5 & 44 &$\mbox{ }$ 0.2 & 38 \\ \hline
$\mbox{ }$ 1.9 & 42 &$*$ 0.9 & 50 &$\mbox{ }$ 0.5 & 44 &$*$ 0.2 & 40 \\ \hline
$\mbox{ }$ 1.7 & 40 &$*$ 0.8 & 38 &$*$ 0.4 & 42 &$*$ 0.2 & 44 \\ \hline
$*$ 1.7 & 46 &$*$ 0.8 & 44 &$*$ 0.4 & 42 &$\mbox{ }$ 0.2 & 42 \\ \hline
$\mbox{ }$ 1.6 & 40 &$*$ 0.8 & 42 &$*$ 0.4 & 36 &$*$ 0.2 & 42 \\ \hline
$\mbox{ }$ 1.5 & 46 &$*$ 0.8 & 32 &$*$ 0.4 & 36 &$*$ 0.1 & 48 \\ \hline
$*$ 1.5 & 40 &$*$ 0.8 & 42 &$*$ 0.4 & 46 &$*$ 0.1 & 46 \\ \hline
$*$ 1.4 & 38 &$\mbox{ }$ 0.7 & 52 &$*$ 0.4 & 44 &$\mbox{ }$ 0.1 & 36 \\ \hline
$*$ 1.4 & 44 &$*$ 0.7 & 40 &$*$ 0.4 & 38 &$*$ 0.1 & 38 \\ \hline
$*$ 1.3 & 38 &$\mbox{ }$ 0.7 & 40 &$\mbox{ }$ 0.4 & 44 &$*$ 0.1 & 26 \\ \hline
$*$ 1.3 & 38 &$*$ 0.7 & 30 &$*$ 0.4 & 38 &$*$ 0.1 & 44 \\ \hline
$\mbox{ }$ 1.3 & 48 &$*$ 0.7 & 46 &$*$ 0.3 & 40 &$\mbox{ }$ 0.1 & 38 \\ \hline
$\mbox{ }$ 1.3 & 40 &$*$ 0.7 & 50 &$*$ 0.3 & 50 &$*$ 0.1 & 40 \\ \hline
$*$ 1.2 & 46 &$*$ 0.7 & 46 &$\mbox{ }$ 0.3 & 36 &$*$ 0.1 & 44 \\ \hline
$\mbox{ }$ 1.2 & 42 &$\mbox{ }$ 0.6 & 48 &$*$ 0.3 & 42 &$*$ 0.1 & 36 \\ \hline
$*$ 1.2 & 38 &$\mbox{ }$ 0.6 & 46 &$*$ 0.3 & 46 &$*$ 0.1 & 50 \\ \hline
$\mbox{ }$ 1.1 & 46 &$\mbox{ }$ 0.6 & 40 &$*$ 0.3 & 40 &$*$ 0.1 & 36 \\ \hline
$*$ 1.1 & 32 &$*$ 0.6 & 46 &$*$ 0.3 & 50 &$*$ 0.1 & 50 \\ \hline
$*$ 1.1 & 32 &$\mbox{ }$ 0.6 & 42 &$*$ 0.3 & 42 &$*$ 0.1 & 46 \\ \hline
$\mbox{ }$ 1.1 & 44 &$\mbox{ }$ 0.6 & 42 &$*$ 0.3 & 42 &$\mbox{ }$ 0.1 & 40 \\ \hline
$*$ 1.1 & 42 &$*$ 0.6 & 34 &$*$ 0.3 & 38 &$*$ 0.1 & 52 \\ \hline
$*$ 1.0 & 40 &$\mbox{ }$ 0.6 & 44 &$\mbox{ }$ 0.3 & 48 &$*$ 0.1 & 50 \\ \hline
      &       &     &       &     &       &$*$ 0.1 & 46 \\ \hline

\end{tabular}
\caption{The 125 structures obtained for sequence W.
The energy
of all these structures is $\Delta E = 0.3$, except for the
ground state that has $\Delta E = 0$, and all states are highly
compact, with a number of contacts $\ge 26$ out of 28 possible.
The first column is the density of hits (probability that
a given structure is hit first) and the second
column is the Hamming distance to the native state.
The ground state is indicated in the table.
The symbol $*$ denotes a compact structure.
}
\label{table2}
\end{center}
\end{table}

We conclude that,
going from the ``protein-like'' sequence A
to sequence C and to the very slow folder
W, one has a spectacular proliferation
of different low-energy conformations
that can be hit starting from a stretched
conformation. It is also worth remarking that the
ground state structure is not the most likely to 
be hit for either sequence C or W.

\begin{figure}
\centerline{\psfig{file=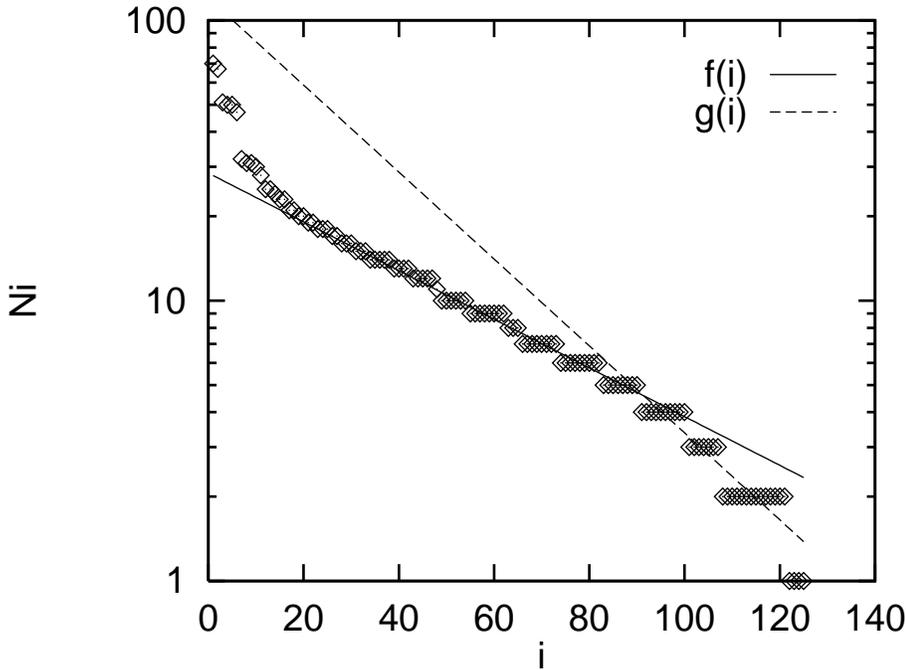,height=8cm}}
\caption{Variations of the number of hits $i$ versus
their rank $i$.
}
\label{Rank}
\end{figure}

In order to determine if any structures with energy
 $\Delta E < \Delta E_{\rm hit}=0.5$ have been missed,
we first notice that for sequence C, the most unlikely
hit has been reached 18 times out of $N=1500$ simulations
(the density of hits for this structure is $1.2 \%$,
as shown in Table \ref{table1}).
This number is much larger than unity, indicating
that all structures of comparable or greater probability 
have been found.
For sequence W, the most unlikely hits have been
reached only once out of the $N=1500$ simulations,
strongly suggesting
that some of the low-energy structures
have not been found. In order to estimate
the number of missed structures for sequence W, 
we have plotted in Fig. \ref{Rank} the
number of times $N_i$ a structure has been
reached, ordered in decreasing order.
The tails have been fitted
to an exponential decay
$f(i) = 28.5 \exp{(-i/50)}$ if $i<90$
and $g(i) = 120 \exp{(-x/28)}$ if
$i>90$.
An estimate of the error in the total number
of structures can be obtained
by extrapolating $g(i)$ and solving
for $g(i)=1$, which leads to $i=134$,
compared to the $125$ different structures
obtained. This indicates that of order 10 
structures with $\Delta E<\Delta E_{\rm hit}=0.5$
were not reached in the $N=1500$ simulations.

We have also compared the states hit in the folding 
simulation to 
the complete set of compact low-energy states
with an energy $\Delta E < \Delta E_{\rm hit}=0.5$ 
(known from the compact-state enumeration \cite{NEC}).
In the case of sequence C, we found that all 12
low-energy states in Table \ref{table1} are indeed
compact, and correspond exactly to all the
compact states with an energy $\Delta E<\Delta E_{\rm hit}=0.5$.
For sequence W, we found 85 compact structures
with an energy $\Delta E < \Delta E_{\rm hit}=0.5$, a number 
considerably smaller than the 125 structures in Table \ref{table2},
which include partially open structures as well. 
Moreover, five of the 85 compact-state 
structures were not hit in our 
simulations. It is very plausible that
these five low-energy compact structures were missed because of
poor statistics for the most unlikely hits; this is consistent with 
the earlier statistical estimate of a total of 10 missed structures. 

The  comparison between the low-energy compact structures and the hit
simulations shows that, for the
average folding sequence C, the 
full low-energy phase space is dynamically accessible,
though some of these low-energy conformations
are more likely to be hit (see Table \ref{table2}).
The proliferation of different low-energy conformations
in the case of the slow folding
sequence W closely matches the large number of 
low-energy compact conformations.
This is natural because the low energy cut-off
$\Delta E_{\rm hit}=0.5$ restricts ``hits'' to compact
or very-close-to compact states.

\subsubsection{``Trapping'' time in the low
energy conformations}
\begin{figure}
\centerline{\psfig{file=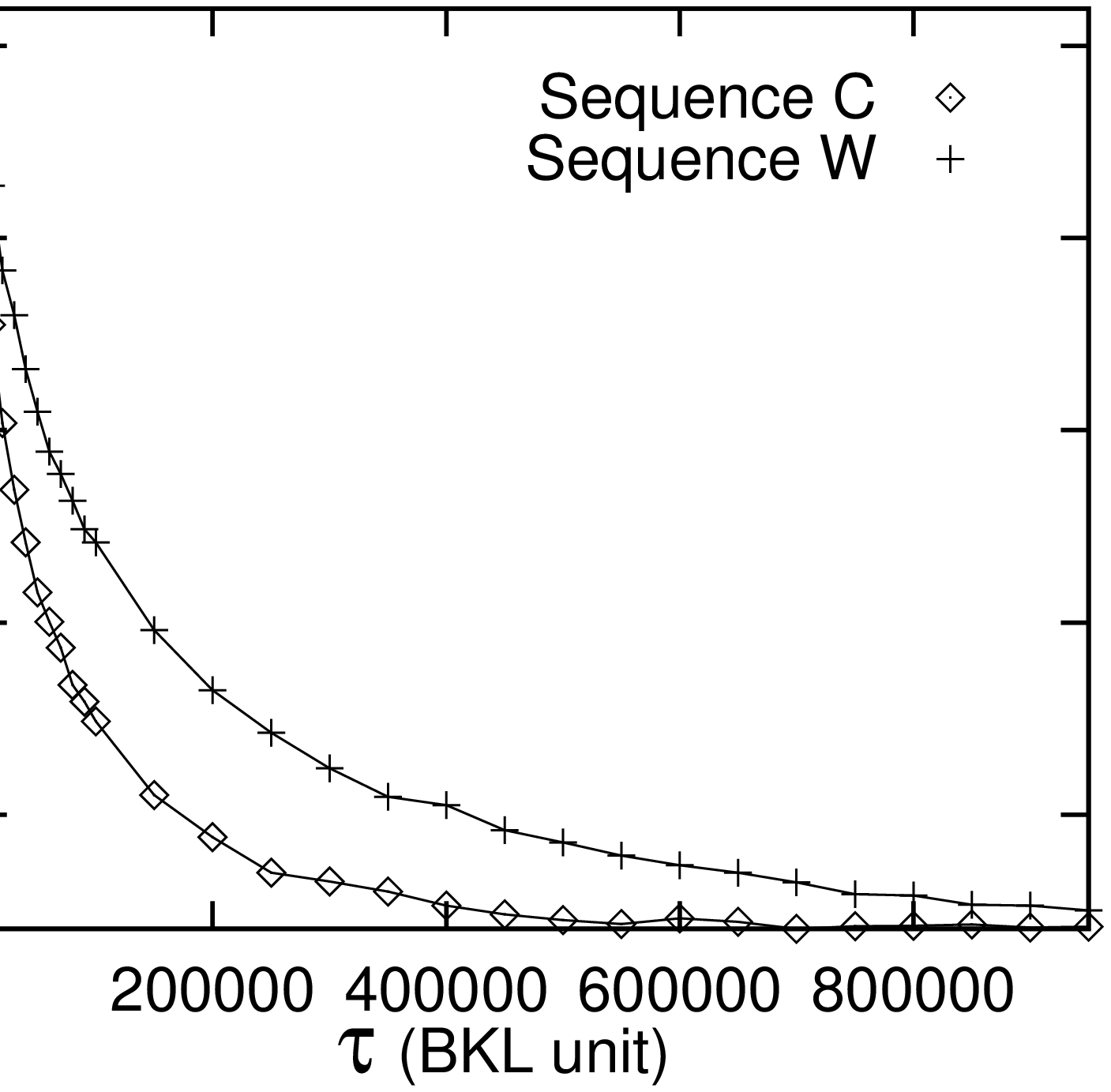,height=8cm}}
\caption{Evolution of the similarity 
$S(t_0,\tau)$ between the conformation at times
$t_0$ and $t_0 + \tau$.
}
\label{Sim}
\end{figure}

We now turn to the question of how long the protein remains
in the first states reached with an energy
$\Delta E < \Delta E_{\rm hit}$. In particular, we would like to
compare the magnitudes of the ``trapping times''
in these low-energy states. To do so, we start from a
stretched conformation and let the dynamics evolve
until a state with energy $\Delta E < \Delta E_{\rm hit}$
is first met at time $t_0$. We then calculate the
similarity $S(t_0,\tau)$ between the conformation at time
$t_0$ and at a later time $t_0 + \tau$.
This quantity is then averaged over $N=1500$ complete simulations.
The results are plotted in Fig. \ref{Sim}
for the sequences C and W, for $\Delta E_{\rm hit} = 0.5$.
As is visible in this figure, the typical 
``trapping time'' in a conformation with an energy
$\Delta E < \Delta E_{\rm hit}$ differs by only a factor
of $2$ or $3$ between the sequences C and W, compared
to a factor of $30$ difference in folding times.
This observation further confirms that the
slow folding of sequence W does not originate
from the trapping in a few ``valleys'' in
phase space with a very long trapping time.
On the contrary, the trapping time of
the average folder C and the very slow
folder W is of the same order of magnitude and
slow folding seems to originate primarily from
the profusion of low-energy conformations for sequence W.

\section{Conclusion}
\label{Conclusion}
Let us now summarize our answers to
the three questions presented in the introduction.
\begin{itemize}
\item (i) We have investigated the relation between
compact-state-spectrum predictions, namely the $\Delta/\Gamma$ criterion,
Eqs. (\ref{Delta}) and (\ref{Gamma}),
and thermodynamic stability, determined from Monte-Carlo
simulations where all the states (not only the compact ones)
are taken into account. We find that for high $\Delta/\Gamma$ the predictions
of the compact-state-spectrum analysis are in good agreement with the
Monte-Carlo simulations. Namely, sequences that have a high
$\Delta/\Gamma$ ratio also have a sharp transition
to the native state conformations (``thermodynamically stable'').
For sequences with low and intermediate 
$\Delta/\Gamma$ ratio, the transition widths vary considerably and are
generally broader. This suggests that low energy open configurations can
be important in determining $\delta$ in this case.  
We find that designability
(defined as the number of sequences that have a given
structure as their nondegenerate ground state) is not in one-to-one
correspondence with thermodynamic stability.  Some sequences with 
highly designable ground states are thermodynamically unstable.
However, we did find that sequences with highly designable ground states
have {\it on average} a large $\Delta/\Gamma$
ratio, and that sequences with poorly designable ground states 
have {\it on average} a small
$\Delta/\Gamma$ ratio.

\item (ii) The folding simulations have shown that sequences with a large
$\Delta/\Gamma$ ratio fold fast. Sequences with a low
$\Delta/\Gamma$ ratio, and therefore sequences
which are thermodynamically unstable, 
may fold slow or fast at the relatively high temperatures
that we investigated. At these
temperatures the protein spends only $10\%$ of the time
near its ground-state configuration.
We have argued, with several examples, that at lower temperatures   
the thermodynamically unstable sequences are likely to
become slow folders whereas the stable ones
are likely to continue to fold fast.

\item (iii) To be ``protein-like'' a sequence must first be thermodynamically stable,
which follows if it has a large $\Delta/\Gamma$ ratio. Second, the sequence
must be mutationally stable, which follows if it has a highly designable 
ground state. We find that the third requirement to be ``protein-like'', namely fast
folding,  does not introduce additional constraints on sequence selection. 
Once a sequence has been
designed to be thermodynamically stable (large $\Delta/\Gamma$)
and stable against mutations (large $N_S$) its folding dynamics is fast.
\end{itemize}

We have also explored the reason for the slow folding
of a sequence with an anomalously large average folding time.
We found
that the ``traping time'' in a given low energy
state is of the same order of magnitude for
both fast and slow sequences,
and cannot
explain the large differences in the
average folding times. Instead, we have shown that
the slow folder visits a large number of
different low-energy states, whereas only a few
low-energy states are visited by the fast folder.
This suggests that slow
folding dynamics originates from a large
number of low energy conformations.

Finally, we make a few remarks to illustrate
the magnitude of structure
and sequence selection.
We started from a set of $2^{27}$ sequences, only
$4.75 \%$ (approximately $6,000,000$) of these having a nondegenerate 
compact ground-state \cite{NEC} structure. The total
number of possible compact structures is $51,704$ \cite{NEC}.
Among all these structures, only $60$ are highly designable
(estimated from the jump of the average compact-state-spectrum gap
as $N_S$ increases above $\simeq 1400$ \cite{NEC}). The $60$
highly designable structures are designed by a total of
$128,320$ sequences, only $15 \%$ of which have $\Delta/\Gamma
> 4.3$ and are thus expected to be ``protein-like'' folders.
In summary, only about $0.01\%$ of the initial
$2^{27}$ sequences satisfy all the  requirements
for ``protein-like'' behavior.

\medskip

\underline{Acknowledgements}: The authors acknowledge
useful discussions with C. Denniston and G. Parisi.
Part of this work was carried out when R.M. was a post-doctoral
fellow at the International School for Advanced Studies
(SISSA) at Trieste.

\end{document}